\documentclass{article}
\usepackage{amssymb}
\usepackage{amsfonts}
\usepackage{amsmath}
\usepackage{lineno}

\input{tcilatex}
\begin{document}

\title{Geometry of Curved Spacetimes Equipped with Torsionful Affinities and
Einstein-Cartan's Theory in Two-Component Spinor Form}
\author{J. G. Cardoso\thanks{%
jorge.cardoso@udesc.br} \\
%EndAName
Department of Mathematics\\
Centre for Technological Sciences-UDESC\\
Joinville 89223-100 SC\\
Brazil.}
\date{ }
\maketitle

\begin{abstract}
The classical world structures borne by spacetimes endowed with torsionful
affinities are reviewed. Subsequently, the definition and symmetry
properties of a typical pair of Witten curvature spinors for such spacetimes
are exhibited along with a comprehensive two-component spinor transcription
of Einstein-Cartan's theory. A full description of the correspondence
principle that interrelates Einstein-Cartan's theory and general relativity
is likewise presented.

\end{abstract}

\section{Introduction}

The most important geometric framework wherein spacetime torsion is
inextricably rooted, comes from the formulation of Einstein-Cartan's theory
[1-3]. This statement relies essentially upon the fact [4-6] that
Einstein-Cartan's theory may be utilized for drawing up alternative
cosmological models which impede the occurrence of the singular
gravitational and cosmological collapses that unavoidably arise in general
relativity [7-12], without imposing any dependence on eventually assignable
symmetries or even on the physical contents of energy-momentum tensors. The
consistency of the construction of the aforesaid torsional cosmological
models stems from the establishment [6] that Einstein-Cartan's equations
admit a two-parameter family of spherically symmetric solutions which supply
a lower bound for the final radius of a gravitationally collapsing cloud of
dust. According to these models, the Universe has expanded in a
non-conformally flat manner from a spherically symmetric state having a
finite radius, without bearing homogeneity insofar as the classical
Friedmann homogeneity property turns out to be lost when torsion is brought
into the spacetime geometry [6]. This latter result has apparently exhibited
a contextual relationship with the work of Ref. [13] which shows that if the
Friedmann cosmological principle such as stated classically is allowed for
inside the realm of Einstein-Cartan's theory, then all the components of the
torsion tensor shall turn out to vanish identically. Moreover, as was
pointed out in Ref. [14], it has provided us with a significant meaning of
the correspondence principle that interrelates Einstein-Cartan's theory and
general relativity.

Very recently, a research program has started whose leading purpose was to
investigate the effects of spacetime torsion, as settled by
Einstein-Cartan's theory, on the generation and propagation of gravitational
waves [15-18]. Through the utilization of an explicit hydrodynamical
modelling for gravitational sources, the so-called post-Newtonian N-body
problem in general relativity has thus been revisited within that torsional
spacetime framework, while some effects coming from radiative losses and
radiation reaction have been systematically described. The achievements
accomplished up till now by this gravitational campaign have then opened up
new cosmological and astrophysical branches.

The characteristic asymmetry borne by the Ricci tensor for any torsionful
world affine connexion always entails the presence of asymmetric
energy-momentum tensors on the right-hand sides of Einstein-Cartan's field
equations. The skew parts of such tensors were originally identified [19-21]
with sources for densities of intrinsic angular momentum of matter that
supposedly generate spacetime torsion locally. It thus became manifest that
spin density of matter plays a physical role in Einstein-Cartan's theory
which is analogous to that played by mass in general relativity.

One of the underlying properties of Einstein-Cartan's theory is that any
spacetime endowed with a torsionful affinity admits a local spinor structure
in much the same way as for the case of generally relativistic spacetimes
[22,23]. This admissibility had supported the construction [24] of a clearly
unique torsional extension of the two-component spinor $\gamma \varepsilon $%
-formalisms of Infeld and van der Waerden for general relativity [25-27].
The construction just referred to has particularly produced a two-component
spinor transcription of Einstein-Cartan's theory [28] which has filled in
the gap concerning the absence from the literature of any acceptable spinor
version of that theory. It had nevertheless been at the outset mainly aimed
at proposing a local description whereby dark energy should be presumptively
looked upon as a torsional cosmic background [29]. Remarkably enough, either
of the torsionful $\gamma \varepsilon $-formalisms affords an irreducible
spinor decomposition for a Riemann tensor, that resembles in form the one
which had been attained much earlier within the framework of general
relativity [30]. It turned out that the physical inner structure of
Einstein-Cartan's theory seems to ascribe a stringent gravitational
curvature-spinor character to the densities of spinning matter which bring
about the local production of spacetime torsion.

In the present paper, we first review the classical world structures that
are borne by spacetimes endowed with torsionful affinities in conjunction
with the formulation of Einstein-Cartan's theory. The settlement and
symmetry properties of typical gravitational curvature spinors for such
spacetimes are shown subsequently along with a comprehensible review of the
above-mentioned spinor version of Einstein-Cartan's theory. A full
explanation of the correspondence principle which takes up Einstein-Cartan's
theory and general relativity, is then given. We believe that our work may
hold an interesting theoretical significance as regards torsional cosmology.
However, we will not work out this feature herein, but we will make some
remarks on it later on.

Our primary motivation for having elaborated upon the spinor formulation of
Einstein-Cartan's theory relies on the belief that it would bring forward
some cosmological pictures which can be ultimately considered as elementary
parts of the theory. Amongst these is the result coming from the spinor form
of Einstein-Cartan's theory according to which it is solely dark matter that
produces spacetime torsion, but the theory has notably suggested that dark
matter should hold a curvature character which is similar to the one
traditionally ascribed to gravitons [28].

The symbolic definitions of world-affine displacements that give rise to
covariant derivatives and Riemann tensors within the torsionful framework,
are similar to those which occur in general relativity [31], and we will not
call upon them at this stage. Also, it will not be strictly necessary to
bring up by this point the spin-affine structures built up in Ref. [24]. The
world and spinor notations used in Ref. [26] will be adopted from the
beginning except that spacetime components will now be labelled by
lower-case Greek letters. We will use the term \textit{trace} to designate
exclusively a world metric trace. A few further conventions will be
explicated in due course.

\section{Torsional Geometry and Einstein-Cartan's Theory}

Within the context of Einstein-Cartan's theory, a spacetime is equipped with
a symmetric metric tensor $g_{\mu \nu }$ of signature $(+---)$ together with
a torsionful covariant derivative operator $\nabla _{\mu }$ that satisfies
the metric compatibility condition\footnote{%
The $\nabla $-operator is taken to possess linearity as well as the
Leibniz-rule property.}%
\begin{equation}
\nabla _{\mu }g_{\lambda \sigma }=0.  \label{addm}
\end{equation}%
The world affine connexion $\Gamma _{\mu \nu \lambda }$ associated to $%
\nabla _{\mu }$ specifies locally an affine displacement in spacetime, and
it is usually split out as%
\begin{equation}
\Gamma _{\mu \nu \lambda }=\widehat{\Gamma }_{\mu \nu \lambda }+T_{\mu \nu
\lambda },  \label{1}
\end{equation}%
where, by definition, $T_{\mu \nu \lambda }=\Gamma _{\lbrack \mu \nu
]\lambda }$ is the torsion tensor of $\nabla _{\mu }$ and%
\begin{equation}
\widehat{\Gamma }_{\mu \nu \lambda }=\Gamma _{(\mu \nu )\lambda }.
\label{1Lin}
\end{equation}%
Hence, $\Gamma _{(\mu \nu )\lambda }$ carries $4\times 10$ independent
components whereas $\Gamma _{\lbrack \mu \nu ]\lambda }{}$ carries $4\times
6 $ such that the number of degrees of freedom of $\Gamma _{\mu \nu \lambda
} $ gets recovered as $40+24.$ In actuality, the tensor character of $\Gamma
_{\lbrack \mu \nu ]\lambda }$ is only related to the symmetry borne by the
inhomogeneous part of the transformation law that describes the behaviour
under the pertinent spacetime mapping group of any world affine connexion
[32,33]. Then, $\Gamma _{(\mu \nu )\lambda }$ does really absorb the
inhomogeneous part that occurs in the transformation law for $\Gamma _{\mu
\nu \lambda },$ whilst $\Gamma _{\lbrack \mu \nu ]\lambda }$ thereby behaves
homogeneously. Owing to this behavioural prescription, we can say that it is 
\textit{not} possible to attain any covariant derivative corresponding to $%
T_{\mu \nu \lambda }$, in contraposition to the case of $\widehat{\Gamma }%
_{\mu \nu \lambda }.$ It follows that the $\nabla $-derivatives of some
purely world vectors $u^{\alpha }$ and $v_{\beta }$ may be written down as 
\begin{equation}
\nabla _{\mu }u^{\lambda }=\text{ }\widehat{\nabla }_{\mu }u^{\lambda
}+T_{\mu \sigma }{}^{\lambda }u^{\sigma },\text{ }\nabla _{\mu }v_{\lambda }=%
\text{ }\widehat{\nabla }_{\mu }v_{\lambda }-T_{\mu \lambda }{}^{\sigma
}v_{\sigma },  \label{3}
\end{equation}%
where $\widehat{\nabla }_{\mu }$ stands for the covariant derivative
operator for $\widehat{\Gamma }_{\mu \nu \lambda }{}$ such that, for example,%
\begin{equation}
\widehat{\nabla }_{\mu }u^{\lambda }=\partial _{\mu }u^{\lambda }+\widehat{%
\Gamma }_{\mu \sigma }{}^{\lambda }u^{\sigma },  \label{3Lin}
\end{equation}%
with\footnote{%
In Eq. (\ref{3Lin}), $\partial _{\mu }$ denotes the partial derivative
operator for some spacetime coordinates $x^{\mu }.$}%
\begin{equation}
\mathcal{A}_{\mu \nu }{}^{\lambda }\doteqdot g^{\lambda \sigma }\mathcal{A}%
_{\mu \nu \sigma }{},  \label{add501}
\end{equation}%
and the kernel letter $\mathcal{A}$ appropriately standing for either $%
\Gamma ,$ $\widehat{\Gamma }$ or $T.$

When acting on a world-spin scalar $f$, the operators $\nabla _{\mu }$ and $%
\widehat{\nabla }_{\mu }$ must agree with each other in the sense that they
should produce common results like $\partial _{\mu }f$. Covariant
derivatives of world tensors of arbitrary valences can be easily obtained
[31] by first performing linear combinations of suitable outer products
between vectors, and then carrying out Leibniz expansions thereof, likewise
implementing the patterns (\ref{3}). It is useful to notice that covariant
derivatives of any tensors may be thought of as involving index displacement
rules. The condition (\ref{addm}) can thus be rewritten as%
\begin{equation}
\widehat{\nabla }_{\lambda }g_{\mu \nu }-2T_{\lambda (\mu \nu )}=0,
\label{4}
\end{equation}%
which right away yields the trace relation%
\begin{equation}
\Gamma _{\mu }=\widehat{\Gamma }_{\mu }+T_{\mu }=\partial _{\mu }\log (-%
\mathfrak{g})^{1/2},  \label{5}
\end{equation}%
with $\Gamma _{\mu }\doteqdot \Gamma _{\mu \sigma }{}^{\sigma }$, for
instance, and $\mathfrak{g}$ denoting the determinant of $g_{\mu \nu }$. It
becomes evident that Eq. (\ref{4}) can account for the secondary metric
condition%
\begin{equation}
\widehat{\nabla }_{\lambda }g_{\mu \nu }=0  \label{addLin1}
\end{equation}%
if and only if the torsion tensor is taken to fulfill throughout spacetime
the subsidiary requirement%
\begin{equation}
T_{\lambda (\mu \nu )}=0.  \label{aff}
\end{equation}%
Putting (\ref{aff}) into effect would, consequently, produce the implications%
\begin{equation*}
T_{\lambda (\mu \nu )}=0\Leftrightarrow \widehat{\nabla }_{\lambda }g_{\mu
\nu }=0\Longrightarrow T_{\lambda \mu \nu }=T_{\lambda \lbrack \mu \nu
]}\Longrightarrow T_{\mu \nu \lambda }=T_{[\mu \nu \lambda ]}\Longrightarrow
T_{\mu }=0,
\end{equation*}%
whence $T_{\mu \nu \lambda }$ will turn out to possess only four independent
components in case either of Eqs. (\ref{addLin1}) and (\ref{aff}) is
actually allowed for.

The crucial meaning of Eq. (\ref{4}) is that any attempt to identify $%
\widehat{\Gamma }_{\mu \nu \lambda }$ with a Riemann-Christoffel connexion
or, in other words, to account for (\ref{addLin1}), requires implementing a
torsion tensor $K_{\mu \nu \lambda }$ which should hold the property%
\begin{equation}
K_{\mu \nu \lambda }=K_{\mu \lbrack \nu \lambda ]}.  \label{add5}
\end{equation}%
Such a tensor would generally supply us at once with the standard $24$
torsional degrees of freedom as we could modify consistently Eq. (\ref{addm}%
) by using in place of (\ref{1}) a contorsion prescription of the type%
\begin{equation}
\breve{A}_{\mu \nu \lambda }=\widehat{\Gamma }_{\mu \nu \lambda }+K_{\mu \nu
\lambda },  \label{af1}
\end{equation}%
while keeping the affine-displacement arrangement of (\ref{add501}). Hence,
we would get the replacement%
\begin{equation}
\nabla _{\mu }g_{\lambda \sigma }=\widehat{\nabla }_{\lambda }g_{\mu \nu
}-2T_{\lambda (\mu \nu )}=0\longmapsto \widehat{\nabla }_{\lambda }g_{\mu
\nu }-2K_{\lambda (\mu \nu )}=\widehat{\nabla }_{\lambda }g_{\mu \nu }=0,
\label{add12}
\end{equation}%
which reinstates (\ref{addLin1}). It turns out that $K{}_{\mu \nu \lambda
}{} $ may be specified in terms of adequate linear combinations of $T_{\mu
\nu \lambda }{}$ which arise out of the utilization of an elegant mechanism
for changing covariant derivative operators [3]. Up to a conventional
overall sign, one gets the definition\footnote{%
We have adopted the definition of the $K$-tensor used in Ref. [34]. The one
used in Ref. [3] differs by an overall minus sign from that of (\ref{A19}).
In any case, Eqs. (\ref{af1}) and (\ref{A19}) yield $\breve{A}_{\mu }=%
\widehat{\Gamma }_{\mu }.$}%
\begin{equation}
K_{\mu \nu \lambda }=\frac{1}{2}(-T_{\mu \nu \lambda }+T_{\nu \lambda \mu
}-T_{\lambda \mu \nu }),  \label{A19}
\end{equation}%
which amounts to the same thing as%
\begin{equation}
K_{\mu \nu \lambda }=\frac{1}{2}(-T_{\mu \nu \lambda }-2T_{\lambda (\mu \nu
)}).  \label{c1}
\end{equation}

It should be stressed that (\ref{A19}) conforms to the skewness-displacement
property%
\begin{equation}
T_{\mu \nu \lambda }=T_{[\mu \nu ]\lambda }\longmapsto K_{\mu \nu \lambda
}=K_{\mu \lbrack \nu \lambda ]},  \label{add1}
\end{equation}%
and satisfies the relations%
\begin{equation}
K_{(\mu \nu )\lambda }=-T_{\lambda (\mu \nu )},\text{ }K_{[\mu \nu ]\lambda
}=-\frac{1}{2}T_{\mu \nu \lambda },\text{ }K_{[\mu \nu \lambda ]}=-\frac{1}{2%
}T_{[\mu \nu \lambda ]},  \label{A1}
\end{equation}%
with the first of which yielding the trace equality%
\begin{equation}
K^{\sigma }{}_{\sigma \mu }=-T_{\mu }.  \label{A5}
\end{equation}%
Therefore, when $\widehat{\Gamma }_{\mu \nu \lambda }{}$ is expressed as a
Riemann-Christoffel connexion, we will get the association%
\begin{equation}
\widehat{\nabla }_{\mu }\longleftrightarrow \frac{1}{2}(2\partial _{(\mu
}g_{\nu )\lambda }-\partial _{\lambda }g_{\mu \nu }),  \label{A2}
\end{equation}%
together with the general prescriptions%
\begin{equation}
\breve{A}_{\mu \nu }{}^{\lambda }=\frac{1}{2}g^{\lambda \sigma }(2\partial
_{(\mu }g_{\nu )\sigma }-\partial _{\sigma }g_{\mu \nu })+K{}_{\mu \nu
}{}^{\lambda }  \label{c4}
\end{equation}%
and%
\begin{equation}
K_{\mu \nu }{}^{\lambda }=\frac{1}{2}(-T_{\mu \nu }{}^{\lambda }+T_{\nu
}{}^{\lambda }{}_{\mu }-T^{\lambda }{}_{\mu \nu }),  \label{c9}
\end{equation}%
which leave the condition (\ref{addm}) invariant as posed by the overall Eq.
(\ref{add12}).

Attention should be drawn to the fact that the number of degrees of freedom
of $K_{\mu \nu \lambda }$ will be reduced to four if $T_{\mu \nu \lambda }$
is required to satisfy Eq. (\ref{aff}). In this particular situation, we
obtain the relations%
\begin{equation}
2K_{[\mu \nu \lambda ]}=-T_{[\mu \nu \lambda ]}=-T_{\mu \nu \lambda
}=2K_{\mu \nu \lambda },\text{ }K^{\sigma }{}_{\sigma \mu }=0,  \label{19}
\end{equation}%
whereas the affinity (\ref{af1}) will carry only $40+4$ degrees of freedom.
Equation (\ref{addm}) could thus be rewritten as\footnote{%
A\ totally skew torsion tensor was used in Ref. [28] to suggest a simpler
procedure for estimating the mass of dark matter.}%
\begin{equation}
g^{\lambda \sigma }\nabla _{\mu }g_{\lambda \sigma }=2K^{\sigma }{}_{\sigma
\mu }=0.  \label{A9}
\end{equation}

Roughly speaking, the Riemann curvature tensor $R_{\mu \nu \lambda \sigma
}{} $ of $\nabla _{\mu }$ carries the information on an invariant difference
between two generally distinct displaced objects which are obtained from
some given world tensor by displacing it along two suitably chosen paths in
spacetime. It appears that the whole information can be extracted from
either of the configurations%
\begin{equation}
\Delta _{\mu \nu }u^{\lambda }=R_{\mu \nu \sigma }{}^{\lambda }u^{\sigma },%
\text{ }\Delta _{\mu \nu }v_{\lambda }=-R_{\mu \nu \lambda }{}^{\sigma
}v_{\sigma },  \label{c5}
\end{equation}%
with the definition%
\begin{equation}
\Delta _{\mu \nu }\doteqdot 2(\nabla _{\lbrack \mu }\nabla _{\nu ]}+T_{\mu
\nu }{}^{\lambda }\nabla _{\lambda }),  \label{c6}
\end{equation}%
which satisfies\footnote{%
By definition, $R_{\mu \nu \lambda }{}^{\sigma }=g^{\sigma \rho }R_{\mu \nu
\lambda \rho }{}.$ The quantity $f$ of (\ref{2}) is a world-spin scalar as
before.}%
\begin{equation}
\Delta _{\mu \nu }f=0.  \label{2}
\end{equation}%
Obviously, the operator $\Delta _{\mu \nu }$ is linear and enjoys the
Leibniz rule property, whence it is legitimate to spell out tensor
expansions like%
\begin{equation}
\Delta _{\mu \nu }s^{\alpha ...\beta }=R_{\mu \nu \tau }{}^{\alpha }s^{\tau
...\beta }{}+\cdots +R_{\mu \nu \tau }{}^{\beta }s^{\alpha ...\tau }{}
\label{e1}
\end{equation}%
and%
\begin{equation}
\Delta _{\mu \nu }w_{\lambda ...\sigma }=-R_{\mu \nu \lambda }{}^{\tau
}w_{\tau ...\sigma }-\cdots -R_{\mu \nu \sigma }{}^{\tau }w_{\lambda ...\tau
}.  \label{e2}
\end{equation}%
So, by invoking the splitting (\ref{1}) together with either of Eqs. (\ref%
{c5}), we are led to the defining expression%
\begin{equation}
R_{\mu \nu \lambda }{}^{\rho }\doteqdot 2(\partial _{\lbrack \mu }\Gamma
_{\nu ]\lambda }{}^{\rho }+\Gamma _{\lbrack \mu \mid \tau \mid }{}^{\rho
}\Gamma _{\nu ]\lambda }{}^{\tau }),  \label{7}
\end{equation}%
which can be reset as%
\begin{equation}
R_{\mu \nu \lambda }{}^{\rho }=\text{\textit{\v{R}}}_{\mu \nu \lambda
}{}^{\rho }+2(\partial _{\lbrack \mu }T_{\nu ]\lambda }{}^{\rho }+T_{[\mu
\mid \tau \mid }{}^{\rho }T_{\nu ]\lambda }{}^{\tau }+T_{[\mu \mid \tau \mid
}{}^{\rho }\widehat{\Gamma }_{\nu ]\lambda }{}^{\tau }+\widehat{\Gamma }%
_{[\mu \mid \tau \mid }{}^{\rho }T_{\nu ]\lambda }{}^{\tau }),  \label{8}
\end{equation}%
where \textit{\v{R}}$_{\mu \nu \lambda }{}^{\rho }$ can be obtained from (%
\ref{7}) by just replacing the kernel letters $R$ and $\Gamma $ with \textit{%
\v{R}} and $\widehat{\Gamma }$, respectively (see Eq. (\ref{p15}) below).

The tensor $R_{\mu \nu \lambda \sigma }{}$ bears skewness in the indices of
each of the pairs $\mu \nu $ and $\lambda \sigma $, but it does not hold the
index-pair symmetry, that is to say,%
\begin{equation}
R_{\mu \nu \lambda \sigma }\neq R_{\lambda \sigma \mu \nu },  \label{T1}
\end{equation}%
whence its Ricci tensor $R_{\mu \nu }\doteqdot R_{\mu \tau \nu }{}^{\tau }$
bears asymmetry.\footnote{%
The Ricci scalar of $\nabla _{\mu }$ is defined by $R=R_{\sigma }{}^{\sigma
}.$} Therefore, $R_{\mu \nu \lambda \sigma }{}$ possesses $36$ independent
components while $R_{\mu \nu }$ possesses $16.$ Some computations show us
that the role of the cyclic identity satisfied by Riemann-Christoffel
curvature tensors, has hereupon to be taken over by%
\begin{equation}
R_{[\mu \nu \lambda ]}{}^{\sigma }-2\nabla _{\lbrack \mu }T_{\nu \lambda
]}{}^{\sigma }+4T_{[\mu \nu }{}^{\tau }T_{\lambda ]\tau }{}^{\sigma }=0,
\label{9}
\end{equation}%
whilst the Bianchi identity should now read%
\begin{equation}
\nabla _{\lbrack \mu }R_{\nu \lambda ]\sigma }{}^{\rho }-2T_{[\mu \nu
}{}^{\tau }R_{\lambda ]\tau \sigma }{}^{\rho }=0.  \label{10}
\end{equation}%
By recalling the dualization schemes constructed in Ref. [3], and making
some index manipulations afterwards, we rewrite Eqs. (\ref{9}) and (\ref{10}%
) as 
\begin{equation}
^{\ast }R^{\lambda }{}_{\mu \nu \lambda }+2\nabla ^{\lambda }{}T_{\lambda
\mu \nu }^{\ast }+4{}T_{\mu }^{\ast }{}^{\lambda \tau }{}T_{\lambda \tau \nu
}=0  \label{11}
\end{equation}%
and\footnote{%
The derivation of Eqs. (\ref{11}) and (\ref{12}) is based on Eq. (3.4.26) of
Ref. [3].}%
\begin{equation}
\nabla ^{\rho }{}^{\ast }R_{\rho \mu \lambda \sigma }+2{}T{}{}_{\mu }^{\ast
}{}^{\rho \tau }R_{\rho \tau \lambda \sigma }=0,  \label{12}
\end{equation}%
where%
\begin{equation}
^{\ast }R{}_{\mu \nu \lambda \sigma }=\frac{1}{2}\sqrt{-\mathfrak{g}}%
\mathfrak{e}_{\mu \nu \rho \tau }R^{\rho \tau }{}_{\lambda \sigma }
\label{nLinLin}
\end{equation}%
and$^{{}}$%
\begin{equation}
T_{\mu \nu }^{\ast }{}^{\lambda }=\frac{1}{2}\sqrt{-\mathfrak{g}}\mathfrak{e}%
_{\mu \nu \rho \tau }T^{\rho \tau \lambda },  \label{nLinLin1}
\end{equation}%
define first left-right duals, with $\mathfrak{e}_{\mu \nu \rho \tau }$
denoting one of the invariant world Levi-Civitta densities. Hence, according
to Eq. (\ref{11}), the contracted dual pattern $^{\ast }R^{\lambda }{}_{\mu
\nu \lambda }$ does not vanish, in contrast to the Riemann-Christoffel case.
By writing out explicitly the expansions of (\ref{10}), and making suitable
index contractions thereafter, we get the equation\footnote{%
Our sign convention for the Ricci tensor comes from the definition $R_{\mu
\nu }\doteqdot R_{\mu \tau \nu }{}^{\tau }$ as shown above.}%
\begin{equation}
\nabla ^{\lambda }R_{\mu \lambda }-\frac{1}{2}\nabla _{\mu }R=2T_{\mu
}{}^{\lambda \sigma }R_{\sigma \lambda }-T_{\sigma \lambda }{}^{\rho }R_{\mu
\rho }{}^{\sigma \lambda }.  \label{15}
\end{equation}

Indeed, the traditional formulation of Einstein-Cartan's theory
intrinsically bears the world geometry characterized by an affine connexion
like the one given as Eq. (\ref{1}). Whereas $T_{\mu \nu }{}^{\lambda }$ is
thus locally related [3] to the spin density of matter $S_{\mu \nu
}{}^{\lambda }$ present in spacetime through%
\begin{equation}
T_{\mu \nu }{}^{\lambda }=-\kappa (S_{\mu \nu }{}^{\lambda }-S_{[\mu }g_{\nu
]}{}^{\lambda }),  \label{e6}
\end{equation}%
with $S_{\mu }\doteqdot S_{\mu \sigma }{}^{\sigma },$ the tensor $g_{\mu \nu
}$ comes into play as a solution to Einstein-Cartan's equations\footnote{%
The quantity $\kappa $ is still identified with Einstein's gravitational
constant of general relativity.}%
\begin{equation}
R_{\mu \nu }-\frac{1}{2}Rg_{\mu \nu }=-\kappa E_{\mu \nu }.  \label{e5}
\end{equation}%
Equation (\ref{e6}) immediately gives the equality%
\begin{equation}
T_{\mu }{}=\frac{\kappa }{2}S_{\mu },  \label{e7}
\end{equation}%
whence we can write the supplementary relation%
\begin{equation}
-\kappa S_{\mu \nu }{}^{\lambda }=T_{\mu \nu }{}^{\lambda }-2T_{[\mu }g_{\nu
]}{}^{\lambda }.  \label{e8}
\end{equation}%
Likewise, working out (\ref{9}) leads us to the somewhat important statement%
\begin{equation}
\nabla _{\lambda }T_{\mu \nu }{}^{\lambda }+2(\nabla _{\lbrack \mu }T_{\nu
]}{}+T_{\mu \nu }{}^{\lambda }T_{\lambda })=\kappa E_{[\mu \nu ]},
\label{e9Lin}
\end{equation}%
which amounts to%
\begin{equation}
(\nabla _{\lambda }+\kappa S_{\lambda })S_{\mu \nu }{}^{\lambda }=-E_{[\mu
\nu ]}.  \label{e12}
\end{equation}%
One can then assert that the skew part of $E_{\mu \nu }$ is a source for $%
S_{\mu \nu }{}^{\lambda },$ and thence also for $T_{\mu \nu }{}^{\lambda }.$
It shall become apparent in the next Section that the importance of Eq. (\ref%
{e9Lin}) is directly correlated to a purely gravitational ascription to the
sources for the densities of spinning matter which produce spacetime torsion
locally, such as mentioned in Section 1.

It is remarked in Ref. [28] that if the trace pattern%
\begin{equation}
T_{\mu }=\nabla _{\mu }\Phi  \label{add3}
\end{equation}%
is taken for granted, with $\Phi $ being a world-spin invariant, then (\ref%
{e8}) and (\ref{e12}) may be fitted together so as to yield the formally
simpler equation%
\begin{equation}
\nabla _{\lambda }T_{\mu \nu }{}^{\lambda }=\kappa E_{[\mu \nu ]},
\label{add15}
\end{equation}%
which would come straightaway from (\ref{e9Lin}) too. Whence, the
implementation of the choice (\ref{add3}) would likewise imply that%
\begin{equation}
\nabla _{\lbrack \mu }T_{\nu ]}{}+T_{\mu \nu }{}^{\lambda }T_{\lambda }=0,%
\text{ }\nabla _{\lbrack \mu }S_{\nu ]}{}=\kappa S_{\mu \nu }{}^{\lambda
}S_{\lambda }  \label{add3Lin}
\end{equation}%
and%
\begin{equation}
\nabla _{\lambda }S_{\mu \nu }{}^{\lambda }+\nabla _{\lbrack \mu }S_{\nu
]}{}=-E_{[\mu \nu ]}.  \label{add3LinLin}
\end{equation}

\section{Einstein-Cartan's Theory in Spinor Form}

In fact, it was the construction of the torsionful version of the classical
Infeld-van der Waerden $\gamma \varepsilon $-formalisms as given in Ref.
[24], that has afforded a coherent two-component spinor transcription of
Einstein-Cartan's theory [28]. In order to carry through the relevant
procedures in a systematic way, one should initially set up the
gravitational curvature spinors of $\nabla _{\mu }$, likewise calling for
the expansions that come from the utilization of the formal
valence-reduction devices provided in Ref. [3]. In this Section, we will
proceed along these lines. Like in the transcription just mentioned, it will
suffice to work out the procedures for the torsional $\varepsilon $%
-formalism, but we will not elaborate upon the geometric spin-density
characterizations borne by it such as brought forth by the work of Ref.
[24]. Thus, any spinor object considered in what follows shall presumably be
viewed as an $\varepsilon $-formalism entity.

The free action of $\Delta _{\mu \nu }$ on some connecting object $\Sigma
_{\lambda }^{AA^{\prime }}$ produces a mixed object $C_{\mu \nu AB}$ that
carries the total information on the spacetime world-spin curvatures. In
Ref. [29], the skew part $C_{\mu \nu \lbrack AB]}$ was taken in regard to
the proposal of a combined description of the cosmic microwave and dark
energy backgrounds. The symmetric part, in turn, bears a purely
gravitational character and amounts to%
\begin{equation}
C_{\mu \nu (AB)}{}=\frac{1}{2}\Sigma _{C^{\prime }A}^{\lambda }{}\Sigma
_{B}^{\sigma C^{\prime }}R_{\mu \nu \lambda \sigma }{}.  \label{s1}
\end{equation}%
We should emphasize that any connecting objects must by definition obey
relations like\footnote{%
The world index of a connecting object has sometimes been displaced just for
an occasional convenience.}%
\begin{equation}
\Sigma _{\mu C^{\prime }}^{(A}\Sigma _{\nu }^{B)C^{\prime }}=\Sigma
_{C^{\prime }[\mu }^{(A}\Sigma _{\nu ]}^{B)C^{\prime }}=\Sigma _{C^{\prime
}[\mu }^{A}\Sigma _{\nu ]}^{BC^{\prime }},  \label{s5}
\end{equation}%
which really ensure the genuineness of the symmetry borne by Eq. (\ref{s1}).

The gravitational curvature spinors of $\nabla _{\mu }$ comprise the
bivector constituents of $C_{\mu \nu (AB)}{},$ and thus occur in the
correspondence%
\begin{equation}
R_{\mu \nu \lambda \sigma }\leftrightarrow (\text{X}_{ABCD},\Xi _{A^{\prime
}B^{\prime }CD}),  \label{s3}
\end{equation}%
whence the X$\Xi $-spinors bear the defining symmetries%
\begin{equation}
\text{X}_{ABCD}=\text{X}_{(AB)(CD)},\text{ }\Xi _{A^{\prime }B^{\prime
}CD}=\Xi _{(A^{\prime }B^{\prime })(CD)}.  \label{s9}
\end{equation}%
One then gets the expressions%
\begin{equation}
R_{AA^{\prime }BB^{\prime }CC^{\prime }DD^{\prime }}=\hspace{-1pt}%
(\varepsilon _{A^{\prime }B^{\prime }}\varepsilon _{C^{\prime }D^{\prime }}%
\text{X}_{ABCD}+\varepsilon _{AB}\varepsilon _{C^{\prime }D^{\prime }}\Xi
_{A^{\prime }B^{\prime }CD})+\text{c.c.}  \label{r1}
\end{equation}%
and\footnote{%
The symbol "c.c." has been taken here as elsewhere to denote an overall
complex conjugate piece.}%
\begin{equation}
^{\ast }R_{AA^{\prime }BB^{\prime }CC^{\prime }DD^{\prime
}}=[(-i)(\varepsilon _{A^{\prime }B^{\prime }}\varepsilon _{C^{\prime
}D^{\prime }}\text{X}_{ABCD}-\varepsilon _{AB}\varepsilon _{C^{\prime
}D^{\prime }}\Xi _{A^{\prime }B^{\prime }CD})]+\text{c.c.}.  \label{r2}
\end{equation}%
Because $R_{\mu \nu \lambda \sigma }$ does not hold the index-pair symmetry,
we also have%
\begin{equation}
\text{X}_{ABCD}\neq \text{X}_{CDAB},\text{ }\Xi _{A^{\prime }B^{\prime
}CD}\neq \Xi _{CDA^{\prime }B^{\prime }},  \label{s15}
\end{equation}%
such that the X$\Xi $-spinors naively recover the number of degrees of
freedom of $R_{\mu \nu \lambda \sigma }$ as $18+18$. Consequently, by
rewriting Eq. (\ref{s1}) as%
\begin{equation}
\frac{1}{2}\Sigma _{CA^{\prime }}^{\lambda }\Sigma _{D}^{\sigma A^{\prime
}}R_{\mu \nu \lambda \sigma }=\Sigma _{M^{\prime }[\mu }^{E}\Sigma _{\nu
]}^{FM^{\prime }}\text{X}_{EFCD}+\Sigma _{M[\mu }^{E^{\prime }}\Sigma _{\nu
]}^{F^{\prime }M}\Xi _{E^{\prime }F^{\prime }CD},  \label{904}
\end{equation}%
likewise utilizing the metric formulae%
\begin{equation}
\Sigma _{AM^{\prime }}^{\mu }\Sigma _{B}^{\nu M^{\prime }}\Sigma _{\lbrack
\mu }^{EA^{\prime }}\Sigma _{\nu ]A^{\prime }}^{F}=-2\varepsilon
^{(E}{}_{A}\varepsilon ^{F)}{}_{B}  \label{m1}
\end{equation}%
and%
\begin{equation}
\Sigma _{AM^{\prime }}^{\mu }\Sigma _{B}^{\nu M^{\prime }}\Sigma _{\lbrack
\mu }^{ME^{\prime }}\Sigma _{\nu ]M}^{F^{\prime }}=\varepsilon
_{(AB)}\varepsilon ^{E^{\prime }F^{\prime }}\equiv 0,  \label{m2}
\end{equation}%
together with the complex conjugates of (\ref{m1}) and (\ref{m2}), one may
pick up the individual X$\Xi $-spinors of (\ref{r1}) in agreement with the
coupling schemes%
\begin{equation}
\frac{1}{2}\Sigma _{AM^{\prime }}^{\mu }\Sigma _{B}^{\nu M^{\prime }}\Sigma
_{CA^{\prime }}^{\lambda }\Sigma _{D}^{\sigma A^{\prime }}R_{\mu \nu \lambda
\sigma }{}=-\Sigma _{AM^{\prime }}^{\mu }\Sigma _{B}^{\nu M^{\prime }}\Sigma
_{\lbrack \mu }^{EA^{\prime }}\Sigma _{\nu ]A^{\prime }}^{F}\text{X}_{EFCD}=2%
\text{X}_{ABCD}  \label{910}
\end{equation}%
and%
\begin{equation}
\frac{1}{2}\Sigma _{MA^{\prime }}^{\mu }\Sigma _{B^{\prime }}^{\nu M}\Sigma
_{CM^{\prime }}^{\lambda }\Sigma _{D}^{\sigma M^{\prime }}R_{\mu \nu \lambda
\sigma }{}=-\Sigma _{MA^{\prime }}^{\mu }\Sigma _{B^{\prime }}^{\nu M}\Sigma
_{\lbrack \mu }^{AE^{\prime }}\Sigma _{\nu ]A}^{F^{\prime }}\Xi _{E^{\prime
}F^{\prime }CD}=2\Xi _{A^{\prime }B^{\prime }CD}.  \label{92}
\end{equation}

With the help of the four-index device [3] 
\begin{align}
\hspace{-0.1cm}\hspace{-0.01cm}\text{X}_{ABCD}\hspace{-0.07cm}& =\hspace{%
-0.07cm}\text{X}_{(ABCD)}-\frac{1}{4}(\varepsilon _{AB}\text{X}%
^{M}{}_{(MCD)}+\varepsilon _{AC}\text{X}^{M}{}_{(MBD)}+\varepsilon _{AD}%
\text{X}^{M}{}_{(MBC)})  \notag \\
& \hspace{-0.07cm}-\frac{1}{3}(\varepsilon _{BC}\text{X}^{M}{}_{A(MD)}+%
\varepsilon _{BD}\text{X}^{M}{}_{A(MC)})-\frac{1}{2}\varepsilon _{CD}\text{X}%
_{AB}{}^{M}{}_{M},  \label{ex}
\end{align}%
we can expand the X-spinor of (\ref{s3}) as%
\begin{equation}
\text{X}_{ABCD}\hspace{-0.07cm}=\hspace{-0.07cm}\Psi _{ABCD}-\varepsilon
_{(A\mid (C}\xi _{D)\mid B)}-\frac{1}{3}\varkappa \varepsilon
_{A(C}\varepsilon _{D)B},  \label{91}
\end{equation}%
with the pieces [24]%
\begin{equation}
\hspace{-0.07cm}\Psi _{ABCD}=\text{X}_{(ABCD)}\hspace{-0.07cm},\text{ }\xi
_{AB}=\text{X}^{M}{}_{(AB)M},\text{ }\varkappa =\text{X}_{LM}{}^{LM}.
\label{92lin}
\end{equation}%
The $\Psi $-spinor of Eq. (\ref{91}) is taken as a typical wave function for
gravitons [35,36], and $\varkappa $ amounts to a complex-valued world-spin
invariant. In Ref. [28], $\xi _{AB}$ was supposed to account for a wave
function for dark matter. Its occurrence in the expansion (\ref{91}) is, in
essence, related to the appropriateness of the first of Eqs. (\ref{s15}).
The complex valuedness of $\varkappa $ enables one to restore readily the
number of complex independent components of X$_{ABCD}$ as $5+3+1$. It
follows that the objects%
\begin{equation}
(\Psi _{ABCD},\xi _{AB},\varkappa ,\Xi _{A^{\prime }B^{\prime }CD})
\label{add2}
\end{equation}%
together determine both $R_{\mu \nu \lambda \sigma }$ and $^{\ast }R_{\mu
\nu \lambda \sigma }$ completely, whence the number of degrees of freedom of 
$R_{\mu \nu \lambda \sigma }$ may be given as $10+6+2+18$. By invoking (\ref%
{r1}) and (\ref{r2}), we can then write out the particular correspondences%
\begin{equation}
R_{\mu \nu }\leftrightarrow R_{AA^{\prime }BB^{\prime }}=\varepsilon
_{AB}\varepsilon _{A^{\prime }B^{\prime }}\func{Re}\varkappa -[(\varepsilon
_{A^{\prime }B^{\prime }}\xi _{AB}+\Xi _{A^{\prime }B^{\prime }AB})+\text{%
c.c.}]  \label{21}
\end{equation}%
and%
\begin{equation}
^{\ast }R^{\lambda }{}_{\mu \lambda \nu }\leftrightarrow R^{CC^{\prime
}}{}_{AA^{\prime }CC^{\prime }BB^{\prime }}=[i(\varepsilon _{A^{\prime
}B^{\prime }}\xi _{AB}-\frac{1}{2}\varepsilon _{AB}\varepsilon _{A^{\prime
}B^{\prime }}\varkappa -\Xi _{A^{\prime }B^{\prime }AB})]+\text{c.c.},
\label{J}
\end{equation}%
which supply us with the parts%
\begin{equation}
R=4\func{Re}\varkappa ,\text{ }^{\ast }R{}_{\mu \nu }{}^{\mu \nu }=4\func{Im}%
\varkappa ,  \label{23}
\end{equation}%
with Eq. (\ref{21}) recovering the number of degrees of freedom of $R_{\mu
\nu }$ as $1+6+9$.

Under certain affine circumstances, the symmetric part of Einstein-Cartan's
equations leads to the limiting case of general relativity. This will be
entertained to a great extent later in the forthcoming Section. Now, we
should instead allow for the skew part%
\begin{equation}
R_{[\mu \nu ]}=-\kappa E_{[\mu \nu ]},  \label{1lin}
\end{equation}%
whose spinor version is, then, constituted by%
\begin{equation}
\varepsilon _{A^{\prime }B^{\prime }}\xi _{AB}+\text{c.c.}=\kappa
(\varepsilon _{A^{\prime }B^{\prime }}\check{E}_{AB}+\text{c.c.}),
\label{add21}
\end{equation}%
where%
\begin{equation}
\check{E}_{AB}=\frac{1}{2}\Sigma _{AC^{\prime }}^{\mu }\Sigma _{B}^{\nu
C^{\prime }}E_{[\mu \nu ]}=\check{E}_{(AB)}  \label{add10}
\end{equation}%
and%
\begin{equation}
\xi _{AB}=\kappa \check{E}_{AB}.  \label{add90}
\end{equation}%
Hence, if we implement the bivector expansions%
\begin{equation}
T_{AA^{\prime }BB^{\prime }}{}^{\mu }=\varepsilon _{A^{\prime }B^{\prime
}}\tau _{AB}{}^{\mu }+\text{c.c.},\text{ }S_{AA^{\prime }BB^{\prime
}}{}^{\mu }=\varepsilon _{A^{\prime }B^{\prime }}\check{S}_{AB}{}^{\mu }+%
\text{c.c.},  \label{e30}
\end{equation}%
after calling for Eqs. (\ref{e6}) and (\ref{e12}), we will get the relation%
\begin{equation}
\tau _{AB}{}^{CC^{\prime }}=-\kappa (\check{S}_{AB}{}^{CC^{\prime }}+\frac{1%
}{2}S_{(A}^{C^{\prime }}\varepsilon _{B)}{}^{C}),  \label{add50}
\end{equation}%
together with $S_{A}^{C^{\prime }}=\Sigma _{A}^{\mu C^{\prime }}S_{\mu }$ and%
\footnote{%
Any $\varepsilon $-spinors and $\Sigma $-objects are covariantly constant
entities.}%
\begin{equation}
(\nabla _{\mu }+\kappa S_{\mu })\check{S}_{AB}{}^{\mu }=-\frac{1}{\kappa }%
\xi _{AB}.  \label{e31}
\end{equation}%
Equation (\ref{e8}) thus gets translated into%
\begin{equation}
-\kappa \check{S}_{AB}{}^{CC^{\prime }}=\tau _{AB}{}^{CC^{\prime
}}+T_{(A}^{C^{\prime }}\varepsilon _{B)}{}^{C},  \label{add51}
\end{equation}%
with $T_{A}^{C^{\prime }}=\Sigma _{A}^{\mu C^{\prime }}T_{\mu }.$ The
dynamical role played by $T_{\mu \nu }{}^{\lambda }$ can be considerably
enhanced if the spinor version of (\ref{add15}) is set up. We have, in
effect,%
\begin{equation}
\nabla _{\mu }\tau _{AB}{}^{\mu }=\xi _{AB},  \label{e32}
\end{equation}%
while the choice (\ref{add3LinLin}) should be transcribed as%
\begin{equation}
\nabla _{\mu }\check{S}_{AB}{}^{\mu }+\frac{1}{2}\nabla _{C^{\prime
}(A}S_{B)}^{C^{\prime }}=-\frac{1}{\kappa }\xi _{AB}.  \label{add5Lin}
\end{equation}

At this point, we can see that it is the combination of Eqs. (\ref{e9Lin}), (%
\ref{add21}) and (\ref{e31}) which tells us that a $\xi $-curvature spinor
must be taken as the only source for spacetime torsion and densities of
intrinsic angular momentum of matter. This result obviously still applies
when the gradient model (\ref{add3}) is taken into account along with Eq. (%
\ref{e32}), and likewise stipulates a precise microscopic character to the
densities of spinning matter that generate $\tau _{AB}{}^{\mu }$ and $\check{%
S}_{AB}{}^{\mu }$.

\section{The Correspondence Principle}

The torsionlessness of the world affine connexions that occur in general
relativity makes it natural to start the description of the correspondence
principle that keeps track of the passage from Einstein-Cartan's theory to
general relativity by taking the limit as $T_{\mu \nu \lambda }$ tends to
zero in Eq. (\ref{1}) and allowing for the splitting (\ref{c4}). This
attitude accounts for the source condition $E_{[\mu \nu ]}=0,$ which implies
that $R_{[\mu \nu ]}=0,$ and the fact that any contorsion tensor will vanish
identically whenever its counterpart torsion does so, in conformity to the
definition (\ref{A19}). In this way, when the torsionless limiting situation
is carried out, $\nabla _{\mu }$ and $\Gamma _{\mu \nu \lambda }$ will
undergo the reductions%
\begin{equation}
\nabla _{\mu }\longmapsto \widehat{\nabla }_{\mu },\text{ }\Gamma _{\mu \nu
\lambda }\longmapsto \widehat{\Gamma }_{\mu \nu \lambda }=\frac{1}{2}%
(2\partial _{(\mu }g_{\nu )\lambda }-\partial _{\lambda }g_{\mu \nu }),
\label{p5}
\end{equation}%
with $g_{\mu \nu }$ having to stand for a solution to Einstein's equations.
The affine association%
\begin{equation}
\widehat{\nabla }_{\mu }\leftrightarrow \widehat{\Gamma }_{\mu \nu \lambda
}{}  \label{J5}
\end{equation}%
should then be specified in such a manner that any allowable covariant
differentiation could involve only a Riemann-Christoffel connexion.

The metric adequacy of (\ref{p5}) can be accomplished by noting that, under
the circumstances being considered, the Riemann tensor of $\nabla _{\mu }$
should be effectively substituted for the one of $\widehat{\nabla }_{\mu },$
namely,%
\begin{equation}
\text{\textit{\v{R}}}_{\mu \nu \lambda }{}^{\rho }\doteqdot 2(\partial
_{\lbrack \mu }\widehat{\Gamma }_{\nu ]\lambda }{}^{\rho }+\widehat{\Gamma }%
_{[\mu \mid \tau \mid }{}^{\rho }\widehat{\Gamma }_{\nu ]\lambda }{}^{\tau
}),  \label{p15}
\end{equation}%
whereas Eqs. (\ref{9}) and (\ref{10}) would have to be replaced with%
\begin{equation}
\text{\textit{\v{R}}}_{[\mu \nu \lambda ]}{}^{\sigma }=0,\text{ }\widehat{%
\nabla }_{[\mu }\text{\textit{\v{R}}}_{\nu \lambda ]\sigma }{}^{\rho }=0
\label{p1}
\end{equation}%
or, equivalently, with%
\begin{equation}
^{\ast }\text{\textit{\v{R}}}^{\lambda }{}_{\mu \nu \lambda }=0,\text{ }%
\widehat{\nabla }^{\rho }{}^{\ast }\text{\textit{\v{R}}}_{\rho \mu \lambda
\sigma }=0.  \label{p2}
\end{equation}%
Like the expression (\ref{7}), \textit{\v{R}}$_{\mu \nu \lambda \sigma }$
satisfies the general property%
\begin{equation}
\text{\textit{\v{R}}}_{\mu \nu \lambda \sigma }=\text{\textit{\v{R}}}_{[\mu
\nu ][\lambda \sigma ]},  \label{add91}
\end{equation}%
whence, by recalling the first of Eqs. (\ref{p1}), we would end up with the
index-pair symmetry%
\begin{equation}
\text{\textit{\v{R}}}_{\mu \nu \lambda \sigma }{}=\text{\textit{\v{R}}}%
_{\lambda \sigma \mu \nu }{},  \label{add100}
\end{equation}%
which leads us to the Ricci tensor of $\widehat{\nabla }_{\mu }$%
\begin{equation}
\text{\textit{\v{R}}}_{\mu \nu }\doteqdot \text{\textit{\v{R}}}_{\mu \sigma
\nu }{}^{\sigma }=\text{\textit{\v{R}}}_{(\mu \nu )}.  \label{add101}
\end{equation}%
Thus, Einstein-Cartan's equations (\ref{e5}) would yield the full Einstein's
equations 
\begin{equation}
\text{\textit{\v{R}}}_{\mu \nu }-\frac{1}{2}\text{\textit{\v{R}}}g_{\mu \nu
}=-\kappa \check{T}_{\mu \nu },  \label{ein1}
\end{equation}%
along with the property $\check{T}_{\mu \nu }=\check{T}_{(\mu \nu )}$ and
the conservation law%
\begin{equation}
2\widehat{\nabla }^{\lambda }\text{\textit{\v{R}}}_{\lambda \mu }{}-\widehat{%
\nabla }_{\mu }\text{\textit{\v{R}}}=0{},  \label{ein2}
\end{equation}%
which constitutes the torsionless version of Eq. (\ref{15}), with a Ricci
scalar \textit{\v{R}} taking over the role of $R.$

Since \textit{\v{R}}$_{\mu \nu \lambda \sigma }{}$ possesses the property
exhibited by Eq. (\ref{add100}), the curvature spinors of $\widehat{\nabla }%
_{\mu }$ must have the symmetries\footnote{%
Without any risk of confusion, we have used the same kernel letters as the
ones of Eq. (\ref{s3}) to denote the curvature spinors of $\widehat{\nabla }%
_{\mu }.$}%
\begin{equation}
\text{X}_{ABCD}=\text{X}_{CDAB},\text{ }\Xi _{A^{\prime }B^{\prime }CD}=\Xi
_{CDA^{\prime }B^{\prime }},  \label{add500}
\end{equation}%
in addition to those given by (\ref{s9}), whilst (\ref{add90}) and the first
of Eqs. (\ref{p2}) will yield%
\begin{equation}
E_{[\mu \nu ]}=0\Longrightarrow \check{E}_{AB}=0\Longrightarrow \xi _{AB}=0,%
\text{ }\func{Im}\varkappa =0.  \label{GR}
\end{equation}%
Then, within the general relativity framework, the X-spinor also has the
property%
\begin{equation}
\text{X}^{M}{}_{(AB)M}=0\Longrightarrow \text{X}_{(ABCD)}=\text{X}_{A(BCD)}=%
\text{X}_{(ABC)D},  \label{j1}
\end{equation}%
while the expansion (\ref{91}) gets simplified to%
\begin{equation}
\text{X}_{ABCD}\hspace{-0.07cm}=\hspace{-0.07cm}\Psi _{ABCD}-\frac{1}{3}%
\varepsilon _{A(C}\varepsilon _{D)B}\func{Re}\varkappa .  \label{j2}
\end{equation}%
The respective $\Xi $-companion thus becomes a Hermitian spinor associated
to a world tensor $\Xi _{\mu \nu }$ via%
\begin{equation}
\Xi _{\mu \nu }=\Sigma _{\mu }^{CA^{\prime }}\Sigma _{\nu }^{DB^{\prime
}}\Xi _{CDA^{\prime }B^{\prime }},  \label{j90}
\end{equation}%
and Eq. (\ref{21}) must be changed to%
\begin{equation}
\text{\textit{\v{R}}}_{AA^{\prime }BB^{\prime }}=\varepsilon
_{AB}\varepsilon _{A^{\prime }B^{\prime }}\func{Re}\varkappa -2\Xi
_{AA^{\prime }BB^{\prime }}.  \label{j30}
\end{equation}%
When combined together, Eqs. (\ref{23}) and (\ref{j30}) produce the
symmetric trace-free definition [3]%
\begin{equation}
-2\Xi _{\mu \nu }=\text{\textit{\v{R}}}_{\mu \nu }-\frac{1}{4}\text{\textit{%
\v{R}}}g_{\mu \nu },  \label{j19}
\end{equation}%
whence Einstein's equations turn out to take the form%
\begin{equation}
\Xi _{\mu \nu }=\frac{\kappa }{2}(\check{T}_{\mu \nu }-\frac{1}{4}\check{T}%
g_{\mu \nu }),  \label{j9}
\end{equation}%
with $\check{T}$ being the trace of $\check{T}_{\mu \nu }.$

The algebraic properties of the X$\Xi $-spinors for general relativity are
described in detail in Refs. [3,27]. There, the expansion for a X-spinor
carries invariants $\Lambda $ and $\chi $ which obey the relations $\func{Re}%
\varkappa =2\chi =6\Lambda .$ The symmetries occurring in Eq. (\ref{add500})
entail reducing the numbers of degrees of freedom of the former
gravitational curvature spinors. For the X$\Xi $-spinors of $\widehat{\nabla 
}_{\mu }$, in effect, we have the prescriptions%
\begin{equation*}
\text{X-spinor}:18-8+1=18-6-1=10+1=11
\end{equation*}%
and%
\begin{equation*}
\Xi \text{-spinor}:18-8-1=4+6-1=10-1=9,
\end{equation*}%
which transparently recover\footnote{%
The prescription $11+9$ was given for the first time in Ref. [30].} the
number of degrees of freedom of \textit{\v{R}}$_{\mu \nu \lambda \sigma }{}$
as $11+9$.

\section{Concluding Remarks}

The explanations regarding the cosmological singularity prevention and
gravitational repulsion, that come straightforwardly from the world form of
Einstein-Cartan's theory, do not require at all the use of the $\gamma
\varepsilon $-formalisms. Some torsional mechanisms have been devised from
this framework [37] which may solve the famous cosmological spatial flatness
and horizon problems without having to call for any outer cosmic
inflationary scenario. One of the central features of the torsional
cosmological models of the Universe based upon Einstein-Cartan's theory, is
that the earliest stages of the cosmic evolution must not have borne
homogeneity, but the classical Friedmann homogeneity property gets
reintroduced into the theoretical framework when the limiting case of
identically vanishing densities of spinning matter is implemented. This
situation brings out a "strong" association between spacetime
torsionlessness and cosmic homogeneity as the contextual occurrence in any
cosmological model of the homogeneity property always demands the absence of
torsion from the spacetime geometry. It seems, then, that the presentation
of the correspondence principle we have exhibited in Section 4 should
incorporate the work of Ref. [13].

From the transcription of Einstein-Cartan's theory replicated in Section 3,
a striking insight into the physical interpretation of the sources for
spacetime torsion has been gained, which could not emerge within any purely
world framework. It may be expected that the treatment of the spatial
flatness and horizon problems could be physically completed on the basis of
the applicability of the torsionful $\gamma \varepsilon $-formalisms to the
models for the birth of the Universe which are derived from
Einstein-Cartan's theory. The theory of dark matter as proposed in Ref. [28]
claims that the sources for dark matter should be prescribed in terms of
well defined couplings between torsion and curvature spinors, while
likewise\ assigning a double physical character to dark matter when the
corresponding wave equations are brought together with Eq.(\ref{e32}).
Emphasis could thus be placed upon the result that the expansion (\ref{91})
is what suggests that gravitons and dark matter were produced together by
the big-bang creation of the Universe, whence we might say that the earliest
cosmic states of very high density of spin matter as predicted by
Einstein-Cartan's theory must have occurred in the absence of conformal
flatness.

We saw that Eqs. (\ref{11}) and (\ref{23}) establish the implication%
\begin{equation*}
\text{torsionlessness of }\nabla _{\mu }\Longrightarrow \text{ reality of }%
\varkappa ,
\end{equation*}%
which means that a necessary condition for a world affinity to bear
torsionlessness is that its $\varkappa $-invariant should bear reality. The
torsionlessness of $\nabla _{\mu },$ on the other hand, surely constitutes a
sufficient condition for $\varkappa $ to bear reality. Such a property can
be helpful to characterize a spacetime geometry from the X-spinor of its $%
\nabla $-operator, in accordance with the relation%
\begin{equation*}
\varkappa =\frac{1}{4}(R+i\text{ }^{\ast }R{}_{\mu \nu }{}^{\mu \nu }).
\end{equation*}

It is shown in Ref. [24] that Eq. (\ref{s1}) possesses the same form as its $%
\gamma $-formalism version. Provided that also the definitions of the metric
spinors and connecting objects for either torsional formalism are formally
the same as the ones for the other formalism, it may be said that the
algebraic description of the spinor pair (\ref{s3}) as well as Eqs. (\ref%
{910}) through (\ref{92lin}), bear the same form in both the formalisms.

ACKNOWLEDGEMENT: I should acknowledge some people for making many
suggestions that have produced a considerable improvement upon the contents
of Sections 1, 4 and 5.

\end{document}